\RequirePackage{fix-cm}
\documentclass{svjour3}                     
\smartqed  
\usepackage{graphicx}
\usepackage{amsmath,graphicx}
\usepackage[ruled,vlined]{algorithm2e}
\usepackage{float}
\usepackage[super]{nth} 
\DeclareMathOperator*{\argmax}{arg\,max}

 \usepackage{subfig}
 \usepackage{hhline}
 \usepackage{tabularx}
 \usepackage{caption}

\usepackage{mwe}
\begin{document}

\title{Good and Bad Boundaries in Ultrasound Compounding: Preserving Anatomic Boundaries While Suppressing Artifacts
}
\titlerunning{Good and Bad Boundaries in Ultrasound Compounding}

\author{Alex Ling Yu Hung         \and
        John Galeotti 
}


\institute{Alex Ling Yu Hung \at
              Biomedical Engineering Department, Carnegie Mellon University, Pittsburgh PA, 15213, USA \\
              \email{lingyuh@andrew.cmu.edu}           
           \and
           John Galeotti \at
              Robotics Institute, Carnegie Mellon University, Pittsburgh PA, 15213, USA \\
              \email{jgaleotti@cmu.edu}
}


\maketitle

\begin{abstract} \hfill \break
\textbf{Purpose} Ultrasound compounding is to combine sonographic information captured from different angles and produce a single image. It is important for multi-view reconstruction, but as of yet there is no consensus on best practices for compounding. 
Current popular methods inevitably suppress or altogether leave out bright or dark regions that are useful, and potentially introduce new artifacts. In this work, we establish a new algorithm to compound the overlapping pixels from different viewpoints in ultrasound.\hfill \break
\textbf{Methods} Inspired by image fusion algorithms and ultrasound confidence, we uniquely leverage Laplacian and Gaussian Pyramids to preserve the maximum boundary contrast without overemphasizing noise, speckles, and other artifacts in the compounded image, while taking the direction of the ultrasound probe into account. Besides, we designed an algorithm that detects the useful boundaries in ultrasound images to further improve the boundary contrast. \hfill \break
\textbf{Results} We evaluate our algorithm by comparing it with previous algorithms both qualitatively and quantitatively, and we show that our approach not only preserves both light and dark details, but also somewhat suppresses noise and artifacts, rather than amplifying them. We also show that our algorithm can improve the performance of downstream tasks like segmentation.\hfill \break
\textbf{Conclusion} Our proposed method that is based on confidence, contrast, and both Gaussian and Laplacian pyramids, appears to be better at preserving contrast at anatomic boundaries while suppressing artifacts than any of the other approaches we tested. This algorithm may have future utility with downstream tasks such as 3D ultrasound volume reconstruction, segmentation, etc.

\keywords{ultrasound image compounding\and ultrasound anatomic boundaries\and ultrasound reconstruction\and Laplacian pyramid}
\end{abstract}

\section{Introduction}
\label{intro}
Even though ultrasound sonography is a low-cost, safe, and fast imaging technique that has been widely used around the globe in clinical diagnosis, surgical monitoring, medical robots etc., there are still some major drawbacks in ultrasound imaging. Due to the nature of how ultrasound images are captured, it can be hard to see the structures that are deep or underneath some highly reflective surfaces \cite{jensen1999linear}. Certain tissues or structures would bounce back or absorb the sound waves, resulting in dark regions underneath.  Such tissues and structures can sometimes produce alterations in ultrasound images which do not represent the actual contents , i.e. artifacts \cite{kremkau1986artifacts}. Moreover, the directionality of ultrasound imaging can make some (parts of) structures difficult to image from certain directions, which may prevent ultrasound images from conveying a complete description of what is going on inside the patient's body. In addition, the directionality may also create confusion for clinicians or medical robots performing downstream tasks. For example, a bullet inside a patient's body would create significant reverberation artifacts that occlude what is underneath. Additionally, when a medical robot inserts a needle into a patient, the reverberation artifacts created by the needle might make the needle tracking algorithm fail or disrupt the identification of the structures of interest \cite{reusz2014needle}. Even though some artifacts have diagnostic significance, which could help clinicians localize certain structures or lesions inside patients' bodies \cite{ahuja1996clinical,baad2017clinical}, the artifacts become less meaningful once the objects of interest are identified. Furthermore, if we preserve the artifacts from different viewpoints, then they could substantially occlude real tissues and the image would be harder to interpret. When there are multiple viewpoints available in ultrasound imaging, we can reconstruct an ultrasound image that represents the underlying structures better while having fewer artifacts.

However, no existing method can do the job perfectly. Relatively simple methods such as averaging the overlapping pixel values from different viewpoints \cite{trobaugh1994three} or taking the maximum of such pixels \cite{lasso2014plus} result in lower dynamic range or additional artifacts in the output images. Other more advanced ultrasound compounding algorithms \cite{gobl2018redefining,hennersperger2015computational} reconstruct the 3D volume of ultrasound using a tensor representation, but both of them still combine the overlapping pixels by simply averaging them or taking the maximum. The method proposed by zu Berge et al. \cite{zu2014orientation} utilizes the per-pixel confidence map proposed by Karamalis et al. \cite{karamalis2012ultrasound} as weights in compounding. While this method does not directly take the average or maximum, it does not take the contrast of the image into account. In the survey paper by Mozafari et al. \cite{mozaffari2017freehand}, a large number of 3D compounding methods are covered, but all of the methods deal with the overlapping pixels from different views by taking the average or maximum.  Since both bright and dark regions contain useful information in ultrasound images, maximizing the contrast in those regions while lowering the intensity of noise and artifacts in other regions is essential in compounding. In all cases, every existing compounding algorithm tends to introduce new artifacts into the volumes and lower the dynamic range. These algorithms can only preserve either dark or bright regions, but in some clinical settings or in computer vision algorithms to guide downstream medical robots, \textit{both dark and bright regions are useful}.

 We do not want to naively take the maximum or average when dealing with overlapping pixels from different views, since doing so would lower the contrast or create new artifacts. Our goal is to clearly differentiate all structures, whether dark or bright, while suppressing artifacts and speckle noise to help with downstream computer vision tasks such as vessel segmentation. To better reconstruct the vessels and detect the bones, unlike prior work, we are less concerned with recovering the most ``accurate" individual pixel values but more concerned with enhancing the images by maximizing the contrast. We focus on preserving patches with the largest contrast, suppressing less-certain high frequency information to prevent piecemeal-stitching artifacts and reduce existing artifacts.  Our most important contributions are:
(1) Use more advanced methods when compounding overlapping pixels between different views instead of directly taking the average or maximum.
(2) Keep the pixels and structures with the higher confidence when compounding.
(3) Preserve the pixels or patches that have the largest local contrast among the overlapping values from different viewpoints.
(4) Identify anatomic boundaries of structures and tissues and treat them differently while compounding.
(5) Use Laplacian pyramid blending \cite{burt1983multiresolution} to remove discrepancy in pixel values from ultrasound images captured in different view points.
(6) Make use of the advantages of different compounding methods in different frequency scales.

\section{Related Work}
As for freehand ultrasound compounding, in 1997, Rohling et al. \cite{rohling1997three} proposed to compound the freehand ultrasound images in the same plane iteratively, using an approach that is based on averaging. Later, the same group used interpolation to reconstruct 3D volumes of non-co-planar freehand ultrasound and still averaged the overlapping pixels \cite{rohling1999comparison}. As mentioned by Rohling et al. \cite{rohling19993d} and Mozaffari et al. \cite{mozaffari2017freehand}, the most common method in freehand compounding is to use interpolation to calculate the missing pixels and use averaging to calculate the overlapping pixels while this might not be the best approach. Grau et al. \cite{grau2005adaptive} came up with a compounding method based on phase information in 2005. Although this method is useful, access to Radio Frequency (RF) data is limited, preventing the algorithms from being widely adopted. Around the same time, Behar et al. \cite{behar2006statistical} showed that averaging the different view worked well if the transducer were set up in a certain way by simulation, but in practice, it would be extremely hard to set up the imaging settings that way. 

In recent years, Karamalis et al. \cite{karamalis2012ultrasound} came forward with a way to calculate physics-inspired confidence values for each ultrasound pixel using a graph representation and random walk \cite{grady2006random}, which Zu Berge et al. \cite{zu2014orientation} used as weights in a weighted average algorithm to compound ultrasound images from different viewpoints. Afterwards, Hung et al. \cite{hung2020ultrasound} proposed a new way to measure the per-pixel confidence based on directed acyclic graphs that can improve the compounding results. Hennersperger et al. \cite{hennersperger2015computational} and Göbl et al. \cite{gobl2018redefining} modeled the 3D reconstruction of ultrasound images based on more complete tensor representations, where they modeled the ultrasound imaging as a sound field. While these two recent papers made great advances in reconstructing ultrasound 3D volumes, they still compound overlapping pixels by averaging or taking the maximum. A review of freehand ultrasound compounding by Mozaffari et al. \cite{mozaffari2017freehand} summarized compounding methods using 2D and 3D transducers. However, few papers talked about how they deal with overlapping pixels, which is what our work mainly focuses on. 

In the case of robot control instead of freehand ultrasound, Virga et al. \cite{virga2018use} modeled the interpolation/inpainting problem as partial differential equations and solved them with a graph-based method purposed by Hennersperger et al. \cite{hennersperger2014quadratic,virga2018use}. They also did image compounding based on the tensor method by  Hennersperger et al. \cite{hennersperger2015computational}. 

Although ultrasound artifacts have barely been directly considered in previous compounding approaches, it has been widely discussed in literature. Reverberation artifacts and shadowing are useful in diagnosis because those artifacts can help clinicians identify highly reflective surfaces and tissues with attenuation coefficients significantly different from normal tissues \cite{hindi2013artifacts}. Reverberation artifacts are most useful in identifying anomalies in lungs \cite{baad2017clinical,soldati2019role}, while it can also be used in thyroid imaging \cite{ahuja1996clinical}. Shadowing could be used in measuring the width of kidneys \cite{dunmire2016use}. However, artifacts and noise could occlude the view of other objects of interests \cite{mohebali2015acoustic} or hurt the performance of other tasks, such as registration \cite{roche2001rigid}, needle tracking \cite{reusz2014needle}, or segmentation \cite{xu2012ultrasound}. In recent years, several learning-based methods have been focusing on identifying artifacts and shadows and using this information to identify other objects \cite{hung2020weakly,meng2019weakly}, but they all need substantial labeling work and a relatively large dataset. Non-learning-based methods to remove the artifacts, either use RF data \cite{tay2011wavelet} or temporal data \cite{win2010identification} or fill in the artifact regions based on neighboring image content within the same image \cite{tay2006transform}. All of these methods make assumptions about what the missing data probably looks like, whereas our approach utilizes multi-view compounding to obtain actual replacement data for the artifact pixel locations.

\section{Methods}
\subsection{Identifying Good Boundaries}
\label{good}
Any sort of averaging between different views in which an object appears either too bright or dark in one view will lower the compounded object's contrast with respect to surrounding pixels. Even though artifacts could be suppressed, the useful structures would also be less differentiated, which is not the optimal approach. Therefore, identifying good anatomic boundaries, and treating them differently than other pixels in compounding, is essential to preserving the dynamic range and contrast of the image. 

Ultrasound transmits sound waves in the axial (e.g. vertical) direction, so sound waves are more likely to be bounced back by horizontal surfaces. Horizontal edges are also more likely to be artifacts, in particular reverberation artifacts \cite{quien2018ultrasound}. The trait of reverberation artifacts is that the true object is at the top with the brightest appearance compared to the lines beneath which are artificial. The distance between the detected edges of reverberation artifacts are usually shorter than other structures. Also, structures in ultrasound images are usually not a single line of pixels: they usually have thickness. Though reverberation artifact segmentation algorithms like \cite{hung2020weakly} could work well in identifying the bad boundaries, labeling images is a very time-consuming task. Besides, the exact contour of the structures in ultrasound images are ambiguous, which can be hard and time-consuming to label as well, so directly using manual labels would be less efficient and it might introduce new artifacts into the images. Therefore, We propose to refine the detected edges based on the appearances of reverberation artifacts. 

First we detect the horizontal boundaries through edge detection algorithms. To detect the actual structures in the ultrasound images instead of the edge of the structure, we calculate the gradient at pixel $(x,y)$ by taking the maximum difference between the current pixel and $\alpha$ pixels beneath, 
\begin{equation}
\frac{\partial I(x,y)}{\partial y}=\max_{j=1,2,..,\alpha}|I(x,y)-I(x,y+j)| 
\end{equation}
where in this paper, we set $\alpha$ to 15.

We then group the pixels that are connected into clusters, such that pixels belonging to the same boundary are in the same cluster. We remove the clusters containing fewer than 50 pixels. After that, we only keep the clusters that do not have a cluster of pixels above itself in $\beta$ pixels. In this paper, $\beta=20$.

A refinement is performed by iterating through the kept clusters and comparing the pixel values against that of the original image. A stack $s$ is maintained, and the pixels in the kept clusters with values greater than $threshold1$ are pushed into it. We pop the pixel $(x,y)$ at the top of the stack and examine the pixels in its 8-neighborhood $(x_n,y_n)$. If $(x_n,y_n)$ has never been examined before and satisfies $I(x_n,y_n)> threshold1$ and at the same time the gradient value is less than $threshold2$, i.e. $|I(x_n,y_n)-I(x,y)|<threshold2$, then we push $(x_n,y_n)$ into the stack $s$. We repeat this procedure until $s$ is empty. We add this step because the boundary detection might not be accurate enough and we can ignore detected boundaries with low pixel values to suppress false positives. In this paper, $threshold1$ and $threshold2$ are set to $30$ and $2$ respectively. The pseudocode for the described algorithm is shown in Algorithm~\ref{Algo_B}. We note that we assigned the values to the parameters based on empirical results. 

\begin{algorithm}[ht]
\KwData{input image $I$}
\KwResult{output boundary mask $B$}
$edges=clustering(denoising(\frac{dI}{dy}))$;\\
$edges=cleanup(edges)$;\\
$mask=zeros(I.shape)$; $B=zeros(I.shape)$;\\
 \For{$edge$ in $edges$}{
 \uIf{$\forall e$ in $edges$ that are far enough from edges underneath}{
 $mask[edge]=1$;
 }
 }

 \For{$[i,j]$ where $mask[i,j]==1$}{
 \uIf{$B[i,j]==0$}{
  stack $s$; \# initialize stack\\
 \uIf{$I[i,j]>threshold1$}{
 $s.push([i,j])$;
 $B[i,j]=1$;
 }
 \While{$s$ is not empty}{
 $[x,y]=s.pop$;\\
 \For{$[ii,jj]$ in the neighborhood of $[x,y]$}{
 \uIf{$I[ii,jj]>threshold1$ and $\mid I[x,y]-I[ii,jj] \mid <threshold2$ and $B[ii,jj]==0$}{
 $s.push([ii,jj])$;
 $B[ii,jj]=1$;
 }
 }
 }
 }
 }
 \caption{Horizontal-edge refinement}
\label{Algo_B}
\end{algorithm}

\subsection{Compounding Algorithm}
Attenuation reduces ultrasound image contrast in deeper regions. Simply taking the maximum, median or mean while compounding \cite{lasso2014plus} further undermines the contrast information, where structure information is stored.
Taking the maximum also would create artifacts by emphasizing non-existent structures resulting from speckle noise in uncertain regions. Although uncertainty-based compounding approach by \cite{zu2014orientation} suppresses the artifacts and noise to some extent, it produces substantially darker images than the originals and lowers the dynamic ranges. Also, taking the maximum retains the bright regions, but some dark regions are also meaningful, so it would make more sense to preserve the patches with the largest local contrast than to simply select the pixels with maximum values. However, directly taking pixels with the largest contrast would lead to neighboring pixels inconsistently alternating between different source images. Besides, the neighbors of a pixel might all be noise, resulting in instability of the algorithm. Taking the maximum contrast might also emphasize the artifacts.

We developed a novel Laplacian-pyramid \cite{burt1983multiresolution} approach to compound the images at different frequency bands and different scales. In this way, we can apply contrast maximization method at certain frequency bands while reconstructing from the pyramid. However, the pixels at extremely large scale in the pyramid represents a patch containing a huge number of pixels in the lower layers, so the contrast in this layer has less anatomical meaning. On the other hand, when the scale is small, the noise in the image would create large local contrast, so maximum weighted contrast might introduce new artifacts into the image. At extremely low and high scales, we thus consider contrast to be less important than intensity confidence measures. Another flaw of directly maximizing the contrast is that the large contrast region might contain artifacts and shadows, so we only maximize the contrast when the overlapping pixels have similar structural confidence values \cite{hung2020ultrasound}, otherwise we use the pixel with the larger structural confidence value in the compounded image, as low  structural confidence value indicates that the pixel belongs to artifacts or shadows. Although some anatomic structures would be removed due to the low confidence values, artifacts and noises would also be removed in the compounded image. The anatomic structures are later compensated for in the later stage of the algorithm. 

Our novel ultrasound compounding method takes ultrasound images from multiple viewpoints and calculates their intensity and structural confidence maps \cite{hung2020ultrasound}, then calculates Laplacian and Gaussian \cite{toet1989image} pyramids of the original images and the Gaussian pyramid of confidence maps. Denote $L_{m,n}$ $GI_{m,n}$ as the n\textsuperscript{th} layer of the Laplacian pyramid and Gaussian pyramid of the m\textsuperscript{th} co-planar ultrasound image respectively, $GC_{m,n}$ $G\Gamma_{m,n}$ as the n\textsuperscript{th} layer of the Gaussian pyramid of the intensity and structural confidence map of m\textsuperscript{th} co-planar ultrasound image respectively, and, $L_k$ as the k\textsuperscript{th} layer of the Laplacian pyramid of the synthetic image. $M$ is the set of viewpoints, with $|M|$ views. Also denote $N(i,j)$ the 8-connected neighborhood of pixel $(i,j)$. Here we combine the weighted maximum contrast and weighted average together. For the k\textsuperscript{th} layer of the pyramid, if the difference across viewpoints between the maximum and minimum structural confidence values $G\Gamma_{m,k}(i,j)$, where $m\in M$, is less than a certain threshold $\gamma$ ($\gamma=0.05$ in this paper), we take the pixel $(i,j)$ with the largest contrast at this scale, since only when there is no artifact at the pixel does taking the largest contrast make sense
\begin{equation}
\widetilde{m}(i,j) = \argmax_{m \in M} \sum_{(a,b)\in N(i,j)}  { \mid GI_{m,k}(a,b)-GI_{m,k}(i,j)\mid}
\label{eq}
\end{equation}
If not, we take the pixel $(i,j)$ with the largest structural confidence at this scale
\begin{equation}
\widetilde{m}(i,j) = \argmax_{m \in M} G\Gamma_{m,k}(i,j)
\end{equation}
Denote the intensity-confidence weighted average at the k\textsuperscript{th} layer of the Laplacian pyramid as $La_k$, 

\begin{equation}
    La_k(i,j)=\frac{\sum_{m=1}^{\mid M \mid} GC_{m,k}(i,j)L_{m,k}(i,j)}{\sum_{m=1}^{\mid M \mid} GC_{m,k}(i,j)}
\end{equation}
Then the k\textsuperscript{th} layer of the Laplacian pyramid of the synthetic image can be calculated as,
\begin{equation}
L_k(i,j)=\phi(k) L_{\widetilde{m}(i,j),k}(i,j)+(1-\phi(k))La_k(i,j)
\end{equation}
where 
\begin{equation}
\phi(k)=\frac{1}{0.4\sqrt{2\pi}}e^{-\frac{1}{2}(\frac{(2k-K-1)^2}{0.16(K-1)^2})}
\end{equation}is a weight function, and K is the number of total layers. This weight function is designed to assign lower weights to contrast maximization and higher weights to intensity-confidence-weighted average in extremely low and high scale.

\begin{figure*}[h]   
\centering
{\includegraphics[width=\textwidth]{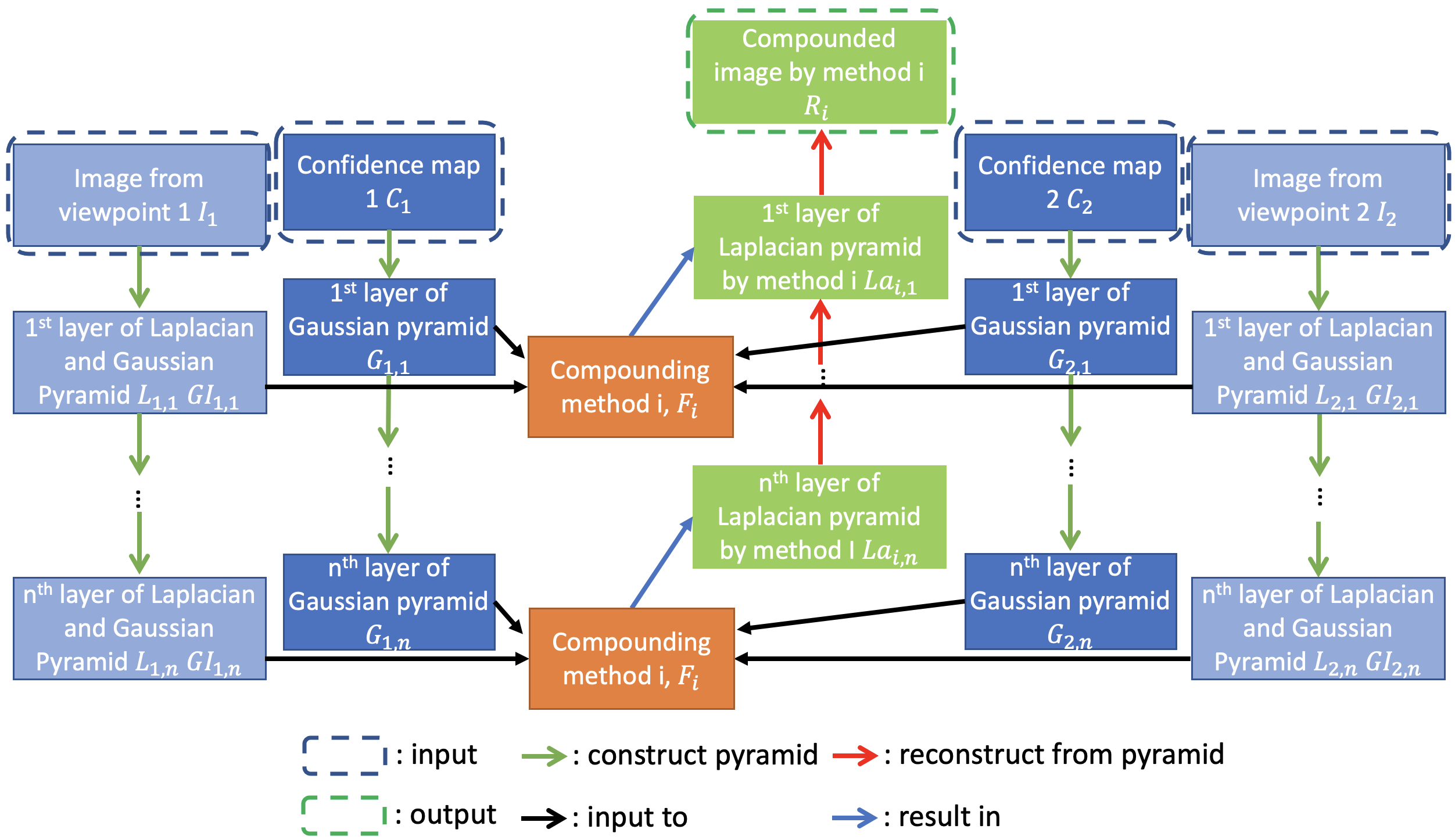}\label{fig:subfig-1}}
\caption[]{Compounding with Laplacian and Gaussian pyramid. The compounding is performed in each layer of the pyramid with the confidence map (intensity confidence or structural confidence) used as some sort of weights. The compounding results are reconstructed from the pyramid of the compounded image.}
\label{subfig-1}
\end{figure*}

The compounding algorithm could be further generalized:

\begin{equation}
L_k(i,j)=\sum_{n=1}^N\phi_n(k)F_n(\{L_{m,k}\}_{m \leq \mid M \mid},\{G_{m,k}\}_{m \leq \mid M \mid})
\end{equation}
where
\begin{equation}
\sum_{n=1}^N\phi_n(k)=1, 
0< k \leq K, 0< n \leq N, 0 \leq \phi_n(k) \leq 1
\end{equation}
$K$ is the total number of layers, $N$ is the total number of compounding methods, $p$ is the total number of viewpoints, $G_{m,k}$ denote any kind of confidence map at layer $k$ from viewpoint $m$, and $F_n$ denote a compounding method. 

We can use any weighting scheme to combine any number of compounding schemes in the Laplacian pyramid based on the application and data.

The algorithm still takes some sort of confidence-based weighted averaging in some layers of the pyramid. During artifact-free contrast maximization, some anatomic boundaries would be removed incorrectly due to lower structural confidence. Therefore, even though this approach works well in preserving contrast and suppressing artifacts, the actual boundaries of structures still tend to get darker. In addition to what we just proposed above, the algorithm we purposed back in section~\ref{good} can also be incorporated. While reconstructing the image from the new Laplacian pyramid after getting the image from the third layer, the good boundaries are detected and values from the original images are taken. For overlapping pixels here, we take the maximum. We apply the same notation as above, and $GB_{m,k}$ is layer $k$ from viewpoint $m$ of the Gaussian pyramid of the boundaries mask $B$ (Gaussian pyramid of algorithm~\ref{Algo_B}'s output).

\begin{equation}
L_3(i,j)=\max(\frac{\sum_{m=1}^{\mid M \mid} {GB_{m,3}(i,j)GI_{m,3}(i,j)}} {\sum_{m=1}^{\mid M \mid} {GB_{m,3}(i,j)}},L_3(i,j))
\end{equation}

This step is done on the third layer of the pyramid since there are still two layers before the final output, so piecemeal-stitching artifacts can still be suppressed. The step isn't done in deeper layers, so that we can still preserve contrast.

\begin{figure*}[h]   
\centering
{\includegraphics[width=\textwidth]{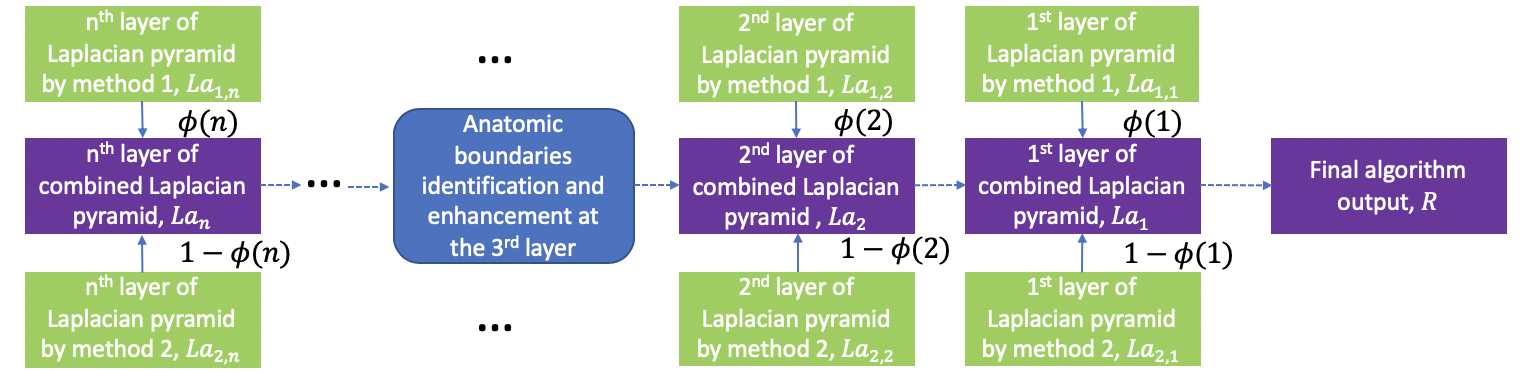}\label{fig:subfig0}}
\caption[]{Pipeline for combining two individual compounding methods and boundaries enhancement. We combine the results from different methods by different weights in each layer of the pyramid. The anatomic boundaries are enhanced at the third layer so that the enhancement does not introduce new artifacts. }
\label{subfig0}
\end{figure*}

\section{Experiments}
\subsection{Data Acquisition}

    The data used in these experiments were gathered from three different sources: a Advanced Medical Technologies anthropomorphic Blue Phantom (blue-gel phantom), an ex-vivo lamb heart, as well as a live pig.  
    
\begin{figure}[h]
     \centering
     \subfloat[][]{\includegraphics[width=0.5\textwidth]{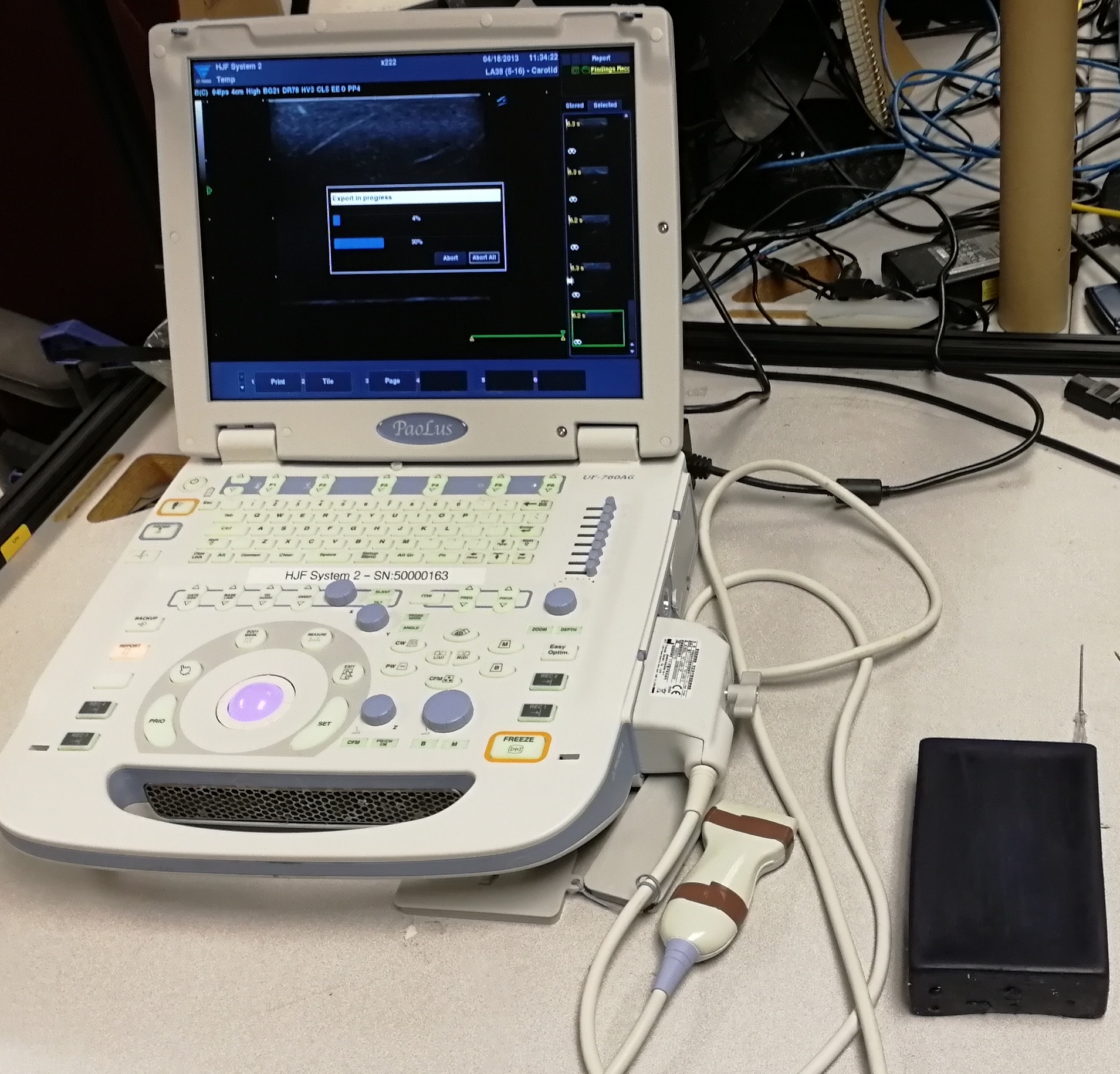}\label{setup1}
     } 
     \subfloat[][]{\includegraphics[width=0.44\textwidth]{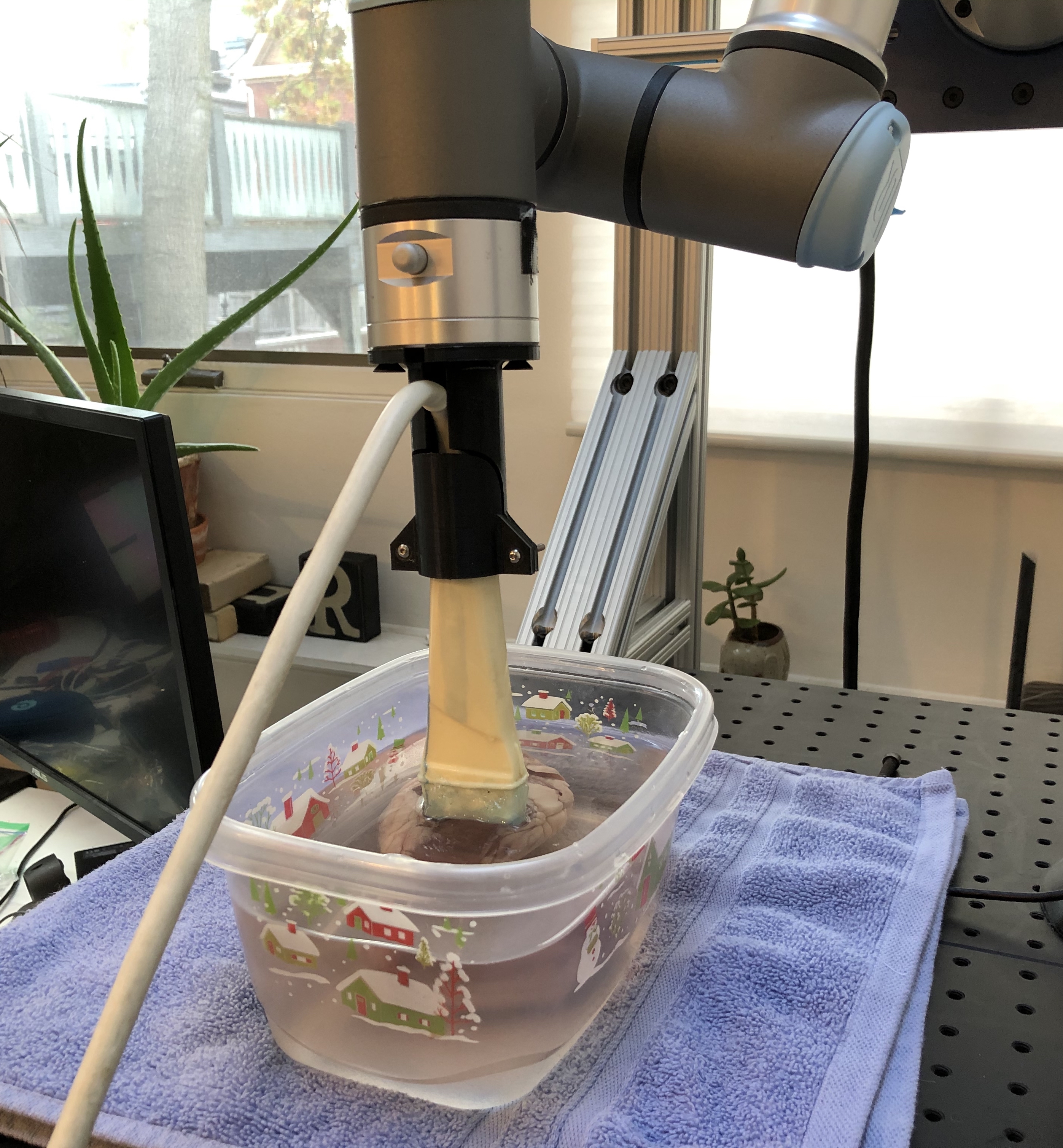}\label{setup2}
     } 
     \caption{(a) The experiment setup for the blue-gel phantom experiment. The square phantom and the needle is shown in the image. We perform the experiment with free-hand imaging since it is easy to make sure the orthogonal views on a square phantom. (b) The experiment setup for the lamb heart experiment, where the lamb heart is situated in a water bath to ensure acoustic coupling. The imaging is done by a robot-controlled high frequency probe.}
     \label{setup}
\end{figure}

    For our initial blue-gel phantom experiments, a UF-760AG Fukuda Denshi diagnostic ultrasound imaging equipment with a linear transducer (51 mm scanning width) set to 12 MHz, a scanning depth of 3 cm and a gain of 21 db is used to scan the surface of the phantom. A needle is rigidly inserted and embedded within the phantom. When scanning the surface, images from two orthogonal viewpoints are collected. As the phantom square it is easy to ensure co-planar orthogonal views with using free-hand imaging without any tracking equipment. The experiment setup is shown in Fig.~\ref{setup1}.
    
    For the ex-vivo experiment, a lamb heart is placed within a water bath in order to insure good acoustic coupling. Using a Diasus High Frequency Ultrasound machine, a 10-22 MHz transducer is rigidly mounted onto a 6 degrees of freedom (dof) Universal Robotics UR3e arm. Using a rough calibration to the ultrasound tip, the 6-dof arm is able to ensure co-planar views of the ex-vivo lamb's heart. The experiment setup is shown in Fig.~\ref{setup2}.
    
    For the in-vivo experiment, a live pig is used as the imaging subject. A UF-760AG Fukuda Denshi diagnostic ultrasound imaging equipment with a linear transducer (51 mm scanning width) set to 12 MHz, a scanning depth of 5 cm and a gain of 21 db is mounted on the end-effector of the UR3e arm and is placed on the desired location manually to get a good view of the vessel \cite{tracir_setup}. Some manual alignments are needed for the arm to be in proper contact with the pig's skin. This pose of the robot is the zero degree view of the vessel. After this, the rotational controller rotates the probe along the probe's tip by the specified angle. For this experiment we cover a range from 20 degree to -25 degree at an interval of 5 deg. The input to the uR3e robot is sent through a custom GUI that is designed to help the users during surgery. The GUI has relevant buttons for the finer control of the robot in the end-effector frame. The GUI also has a window that displays the ultrasound image in real-time which helps in guiding the ultrasound probe.

\subsection{Qualitative Evaluation}
\begin{figure}[h]   
\centering
{\includegraphics[width=\textwidth]{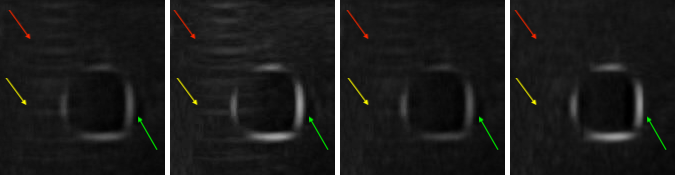}}
\caption[]{Compounded patches left to right: average \cite{trobaugh1994three}, maximum \cite{lasso2014plus}, uncertainty-based fusion \cite{zu2014orientation}, and our algorithm, where the green arrows indicate the vessel walls while the red and yellow arrows indicate the artifacts. As is shown in the figure that our result preserves the brightness of the vessel boundaries and suppresses the artifacts at the same time, while other methods fail to do so.}
\label{comp}
\end{figure}

We visually compare the results of our method against average \cite{trobaugh1994three}, maximum \cite{lasso2014plus}, and uncertainty-based fusion \cite{zu2014orientation}.  As is shown in Fig.~\ref{comp}, our algorithm has the best result in suppressing artifacts, and at the same time, the brightness of the boundaries (green arrows) from our algorithm is similar to that of taking maximum \cite{lasso2014plus}. Our method also preserves a lot more contrast since other parts of the patch are darker in comparison to our bright boundaries, whereas the boundaries from the other two compounding algorithms are darker and therefor less contrasting with the dark interior. Our algorithm also completely suppresses the reverberation artifacts in the regions that the red and yellow arrows point to, while the results from other algorithms all preserve undesirable aspects of artifacts. 

\begin{figure*}[h]   
\centering
{\includegraphics[width=\textwidth]{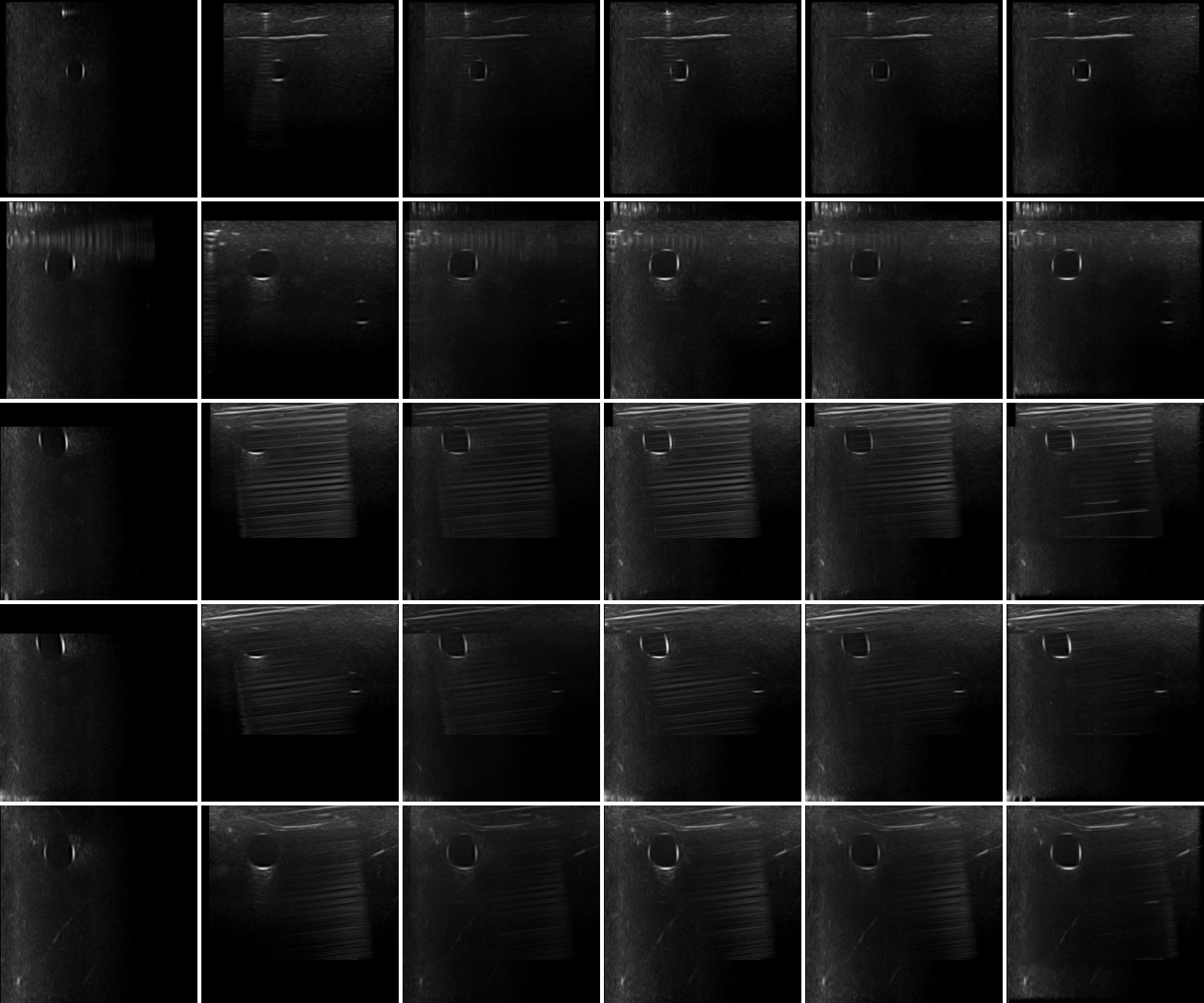}}
\caption[]{Results on the phantom with a needle inserted in it. The left two columns are the two input images
(phantom images were acquired orthogonally within plane, where the imaging direction of the first and second column are from left to right and from top to bottom respectively).
The right four columns from left to right are results from average \cite{trobaugh1994three}, maximum \cite{lasso2014plus}, uncertainty-based fusion \cite{zu2014orientation}, and our algorithm. On the phantom examples it is clear that our method best preserves bright and dark anatomy while suppressing artifacts.}
\label{comp2}
\end{figure*}

To compare our results against other existing compounding algorithms (average \cite{trobaugh1994three}, maximum \cite{lasso2014plus}, and uncertainty-based fusion \cite{zu2014orientation}), we select 5 examples of results on the anthropomorphic phantom, which is shown in Fig.~\ref{comp2}. In the first row, our algorithm almost completely removes the reverberation artifacts in the synthesized image and at the same time preserves the contrast in the images. In other phantom examples, our algorithm is also the best in removing the reverberation artifacts and shadows the vessel walls cast, while preserving the brightness of the vessel walls, needles and other structures in the images. Our algorithm preserves the``good boundaries" that represent the anatomic boundaries while suppressing boundaries that are not real.

\begin{figure*}[h]   
\centering
{\includegraphics[width=\textwidth]{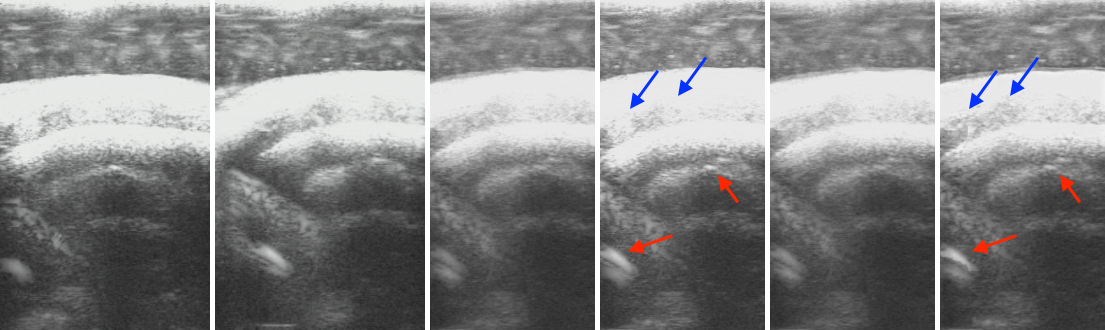}}
\caption[]{Results on the lamb heart ultrasound images. From left to right: two input images, compounded results (only overlapped regions are shown) by average \cite{trobaugh1994three}, maximum \cite{lasso2014plus}, uncertainty-based fusion \cite{zu2014orientation}, and our algorithm. Maximum\cite{lasso2014plus} and ours the only methods that are able to preserve the bright boundaries at the red arrows, but the maximum is not able to preserve the contrast at the blue arrows like ours does.}
\label{lamb}
\end{figure*}

Besides, we also test our algorithm on real tissue images. The comparison between our algorithm and other existing algorithms on the lamb heart is shown in Fig.~\ref{lamb}, where only maximum \cite{lasso2014plus} and our algorithm preserve the contrast at the red arrows, whereas the results by other algorithms are darker in that patch. However maximum \cite{lasso2014plus} fails to preserve the contrast at the blue arrows, while our method keeps the contrast at both red and blue arrows. It also shows that even on highly noisy data, our algorithm also has decent performance.

\begin{figure*}[h]   
\centering
{\includegraphics[width=\textwidth]{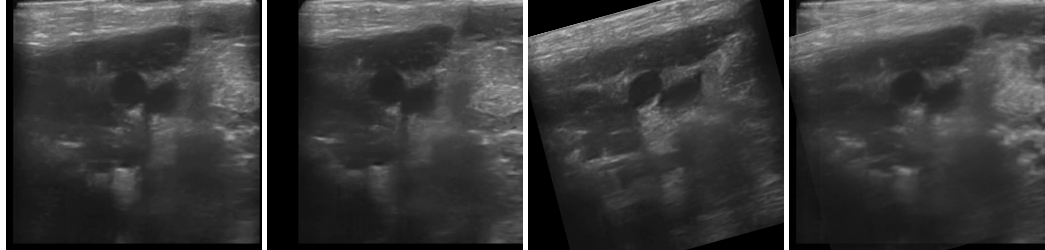}}
\caption[]{The compounding result of three live-pig images. The left three images are the input images while the right image is the result. In the result, the vessel and the structure on the right is successfully preserved while the shadows cast by the vessel become less significant.}
\label{pig}
\end{figure*}

We further demonstrate that our method is able to handle images from more than two viewpoints, i.e. situation where $|M|>2$. In this experiment, we utilize the live-pig data and instead of using the structural confidence, we use a simple contrast maximization (equivalent to the case when all structural confidence at corresponding pixels are equal), due to the difficulty in getting the reference image for the structural confidence. The result is shown in Fig.~\ref{pig}, where the change in probe position between the first two images only consists of translation, and when moving the probe to the third location it also involves rotation. 

\begin{figure}[h]   
\centering
{\includegraphics[width=\textwidth]{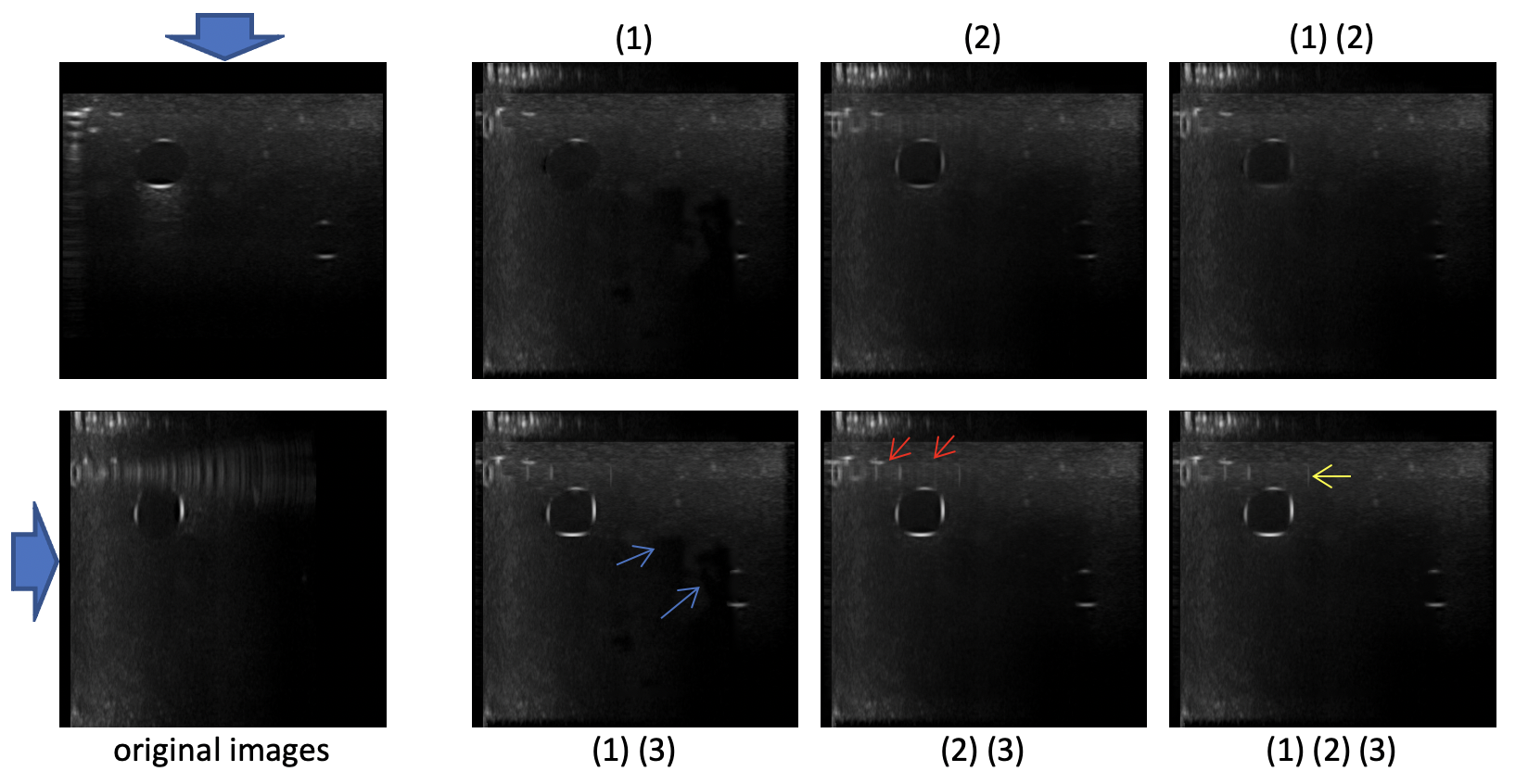}}
\caption[]{Results showing the effect of different parts of the algorithm. The left column consists of the original images with arrows indicating the imaging direction. The right 6 images are compounding results where the numbers above or under the images indicate which part(s) of the algorithm is (are) used to constructed the compounded images. Note that the correspondence of the numbers are (1) structural confidence based artifact-free contrast maximization, (2) intensity confidence based weighted averaging, (3) edge enhancement.}
\label{aba}
\end{figure}

 We also would like to show how each component of our algorithm contribute to the final output. Our proposed algorithm mainly consists of three parts: (1) structural confidence based artifact-free contrast maximization, (2) intensity confidence based weighted averaging, (3) edge enhancement. Shown in Fig.~\ref{aba}, structural confidence based artifact-free contrast maximization (1) removes the reverberation artifacts and shadows decently well, and preserves some contrast in the images, but some parts of the vessel boundaries are removed as well and creates some unnatural holes in the image. Intensity confidence based weighted averaging (2) preserves the vessel boundaries but not as bright as before, and it also removes the reverberation artifacts but also not as good as structural confidence based artifact-free contrast maximization (1). Edge enhancement (3) clearly enhances the boundaries but at the same time slightly enhances a small portion of the reverberation artifacts (yellow arrow) as well. Generally, the final output ((1)(2)(3)) leverages the different components of the image, having less reverberation artifacts than the result by using only (2) and (3) (red arrow), while having no irregular holes like the result by only (1) and (3) (blue arrow). Depending on different application, we can adjust the weights $\phi(k)$ and how we utilize the detected good boundaries, to compound the images in the way we want. 

\subsection{Quantitative Evaluation}
We continue to compare our results with average (avg) \cite{trobaugh1994three}, maximum (max) \cite{lasso2014plus}, and uncertainty-based fusion (UBF) \cite{zu2014orientation}, as well as the original images. The challenges to evaluate the results are (1) there are no ground truth images that show what the compounded images should look like, (2) our algorithm is designed to maximize the contrast near boundaries and suppress the artifacts, so the exact pixel values do not matter, so manually labeled binary masks where anatomic boundaries are 1 and other pixels are 0 would not work as some naive ground truths. Besides, since the majority of the pixels would be 0 in those naively labeled images, the peak signal to noise ratio (PSNR) with such images as ground truth would be a lot larger if the images are dark compared with images with larger visual contrast. 

\begin{figure}[h]   
\centering
{\includegraphics[width=\textwidth]{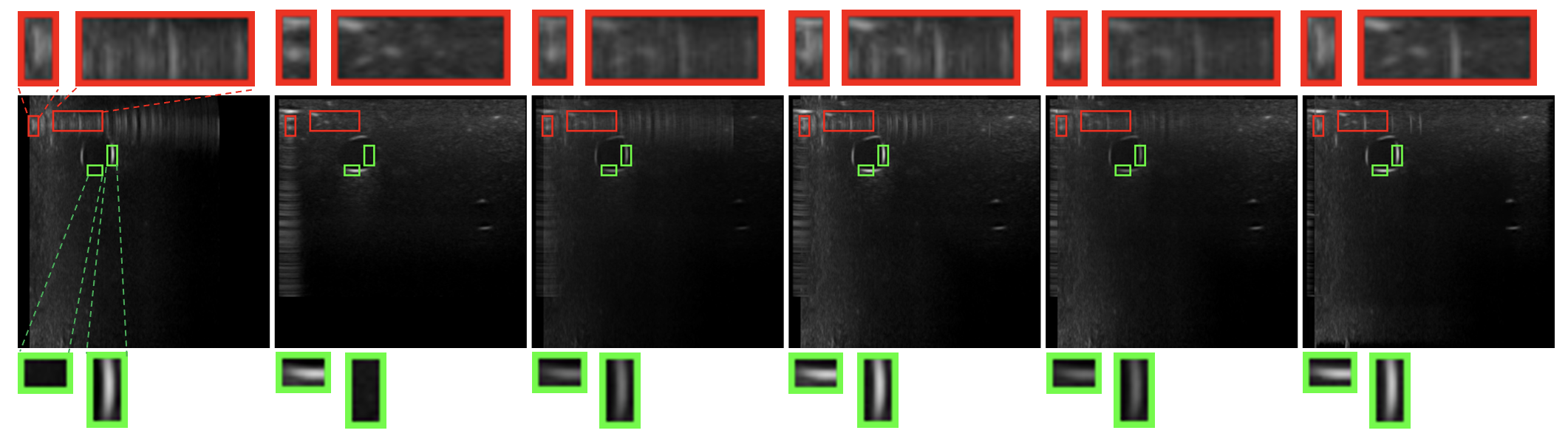}}
\caption[]{An example of boundary signal and artifact patches. From left to right are two original images that are orthogonal, followed by results from average \cite{trobaugh1994three}, maximum \cite{lasso2014plus}, uncertainty-based fusion \cite{zu2014orientation}, and our algorithm. The green and red boxes are examples of boundary and artifact patches. }
\label{ex}
\end{figure}

To show that our method generates images with better quality, we propose to use our variance-based metric. We separately evaluate image patches containing artifacts, which should have low contrast; and patches containing boundaries, which should have high contrast. For the patches with artifacts, we evaluate the algorithms based on the ratio between the variance of the patch and the variance of the whole image (denoted as variance ratio), as well as the ratio between the mean of the patch and the mean of the whole image (denoted as mean ratio). The patches with the artifacts should have lower variance and a similar mean compared with the whole image, since artifacts are supposed to be suppressed. As for patches with real boundary signals, we only care about the contrast, so our metric is the variance ratio. We want the variance in the patches with boundary signals to be much larger than the variance of the whole image. We compute the average mean ratio (AMR), average variance ratio (AVR) on 27 signal patches and 23 artifact patches. These patches are cropped from the same position in every image to keep the comparison fair, and examples of the patches are shown in Fig.~\ref{ex}, where the green boxes are the anatomic boundary signal patches and the red boxes are the artifact patches. The results are listed in Table ~\ref{tab1}. Our method outperforms other algorithms in suppressing artifacts. As for real boundary signals, our method appears superior to all the other methods.

\begin{table}[h]
\centering
\caption{Evaluation by Mean and Variance}
\label{tab1}
\begin{tabular}{|c|c|c|c|c|c|c|c|}
\hline
{} & {} &view1 &view2&avg & max &UBF & ours\\ 
\hline
{artifacts} & \begin{tabular}{c}
                   AMR\\
                   AVR\\
                 \end{tabular}
                 & \begin{tabular}{@{}l@{}}
                   1.434\\
                   0.109\\
                 \end{tabular}
                 & \begin{tabular}{@{}l@{}}
                   1.996\\
                   0.224\\
                 \end{tabular}
                 & \begin{tabular}{@{}l@{}}
                   1.757\\
                   0.204\\
                 \end{tabular}
                 & \begin{tabular}{@{}l@{}}
                   1.433\\
                   0.234\\
                 \end{tabular}
                 & \begin{tabular}{@{}l@{}}
                   1.277\\
                    0.134\\
                 \end{tabular}
                 & \begin{tabular}{@{}l@{}}
                   \bfseries 1.206\\
                   \bfseries 0.048\\
                 \end{tabular}\\
\hline
{boundaries} & \begin{tabular}{c}
                   AVR\\
                 \end{tabular}
                 & \begin{tabular}{@{}l@{}}
                   3.609
                 \end{tabular}
                 & \begin{tabular}{@{}l@{}}
                   2.172
                 \end{tabular}
                 & \begin{tabular}{@{}l@{}}
                   3.007
                 \end{tabular}
                 & \begin{tabular}{@{}l@{}}
                   3.612
                 \end{tabular}
                 & \begin{tabular}{@{}l@{}}
                   1.696
                 \end{tabular}
                 & \begin{tabular}{@{}l@{}}
                   \bfseries 3.876
                 \end{tabular}\\
\hline
\end{tabular}

\end{table}

\begin{figure*}[h]   
\centering
{\includegraphics[width=\textwidth]{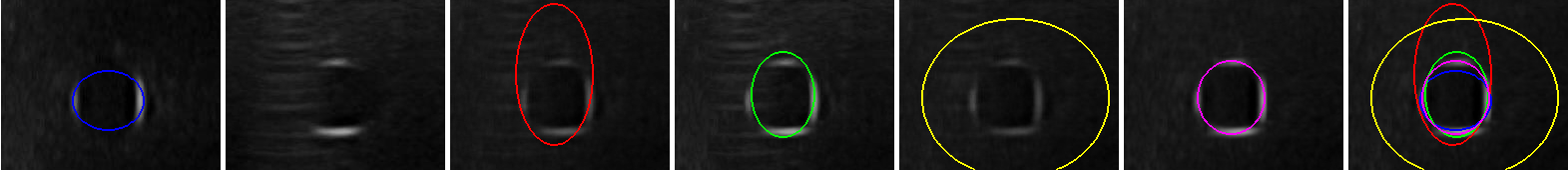}}
\caption[]{The result of vessel segmentation. First two images: two original images (the algorithm is not able to fit an ellipse on the second image). Following four images: results by average \cite{trobaugh1994three}, maximum \cite{lasso2014plus}, uncertainty-based fusion \cite{zu2014orientation}, and our algorithm. The last image: Segmentation results overlaying on the compounded image synthesized by our algorithm. It can be seen that the segmentation on the first original image is flatter because of the missing top and down boundaries, while the segmentation on the result by maximum is affected by the reverberation artifact at the top. Other segmentation results are clearly off, while the segmentation algorithm fits the vessel boundaries very well on our result. }
\label{seg}
\end{figure*}

Additionally, we also compare our results with the previous ones by performing vessel segmentation on the compounded images. We manually selected 6 patches containing vessels with annotated the vessel boundaries as ground truth. We perform a naive segmentation as the following. We first apply Otsu thresholding \cite{otsu1979threshold} to each compounded patch which contains blood vessels to automatically separate the vessel boundaries from the background. We then fit an ellipse to the separated boundary points by the first step to segment the vessel. Table~\ref{tab2} shows dice coefficients \cite{zou2004statistical} comparing each method against ground truth where ours have the best performance. Fig.~\ref{seg} shows an example of the segmentation result. This simple adaptive segmentation performs the best on our compounding results. Since Otsu thresholding is purely pixel intensity based thresholding without considering other information, it is somewhat sensitive to the intensity of noise in the image. Therefore, better segmentation results show that our method is better than the prior algorithms at preserving the vessel walls while suppressing noise and artifacts.

\begin{table}[h]
\centering
\caption{Evaluation by Segmentation}\label{tab2}
\begin{tabular}{|c|c|c|c|c|c|c|}
\hline
 {} &view1 &view2&avg & max &UBF & ours\\ 
\hline
{Dice Coefficient} &0.654&0.575& {0.673} & {0.719} &{0.594} &{\bfseries 0.737}\\
\hline
\end{tabular}
\end{table}

\section{Conclusion}
In this work, we present a new ultrasound compounding method based on ultrasound per-pixel confidence, contrast, and both Gaussian and Laplacian pyramids, taking into account the direction of ultrasound propagation. Our approach appears better at preserving contrast at anatomic boundaries while suppressing artifacts than any of the other compounding approaches we tested. Our method is especially useful in compounding problems where the images are severely corrupted by noise or artifacts and there is substantial information contained in the dark regions in the images. We hope our method could become a benchmark for ultrasound compounding and inspire others to build upon our work. Potential future work includes 3D volume reconstruction, needle tracking and segmentation, artifact identification and removal, etc.

\begin{acknowledgements}
We thank our collaborators at the University of Pittsburgh, Triton Microsystems, Inc., Sonivate Medical, URSUS Medical LLC, and Accipiter Systems, Inc. We also thank Evan Harber, Nico Zevallos, Abhimanyu, Wanwen Chen and Prateek Gaddigoudar from Carnegie Mellon University for gathering the data and reviewing the paper. This is a post-peer-review, pre-copyedit version of an article
published in [insert journal title]. The final authenticated version is available online at:
https://doi.org/10.1007/s11548-021-02464-4.
\end{acknowledgements}

%
 \section*{Declaration}
\textbf{Funding} This work was sponsored in part by a PITA grant from the state of Pennsylvania DCED C000072473, and by US Army Medical contracts W81XWH-19-C0083, W81XWH-19-C0101, W81XWH-19-C-002.\\
\textbf{Conflict of interest} Galeotti serves on the advisory board for Activ Surgical, Inc., and he is a Founder and Director for Elio AI, Inc.\\
\textbf{Ethical approval} All applicable international, national, and/or institutional guidelines for the care and use of animals were followed.\\
\textbf{Informed consent} Informed consent was obtained from all individual participants included in the study.

\bibliographystyle{spmpsci}      

\bibliography{a.bib}   

\end{document}